# DEVELOPING THE APPLICATION FOR LEARNING PLACE VALUE

Rawiphon Charunphankaseam[1]

[1]Faculty of Mass Communication Technology, Rajamangala University of  Technology Phra Nakhon, Bangkok, Thailand

*ABSTRACT*

*The objectives of this research were 1) to develop the application for learning place value, 2) to determine the efficiency of developed application, 3) to compare academic achievement to compare academic achievement between pre-lesson and post-lesson of students who learned with the developed application about place value, 4) to compare academic achievement between students who learned place value by the developed application and through traditional method during post-lesson, 5) to compare retention of academic achievement of application for learning place value after 2 weeks, and 6) to find satisfactory of student who learned the application for learning place value. The sample group selected through purposive sampling was 5 content specialists and 400 pratomsuksa 1 students. The result of the study showed the qualitative assessment of the application for learning place value that was developed had overall highest level of efficiency ( $\bar{x}$= 4.73, S.D. =0.47) of the application for learning place value. Pratomsuksa 1 students showed with overall qualitative assessment equivalent to 89.07/91.84 that was higher than anticipated as 80/80.  The academic achievement between pre-lesson and post-lesson by the application for learning place value appeared an average of 31.71 in pre lesson and an average of 53.49 in post-lesson. The comparison of the academic achievement during pre-lesson and post-lesson of when students learned place value with the developed application showed 0.5 statistical significance. The academic achievement between the students with the application for learning place value an average of 53.49 during and an average of 41.33 during post-lesson. From the compare academic achievement between students who learned place value by the developed application and through traditional method during post-lesson, the student learning with the application for learning place value had higher academic achievement than those learning in the traditional method at 0.5 statistical significance. The retention of the students learning in post-lesson with the application for learning place value after 2 weeks did not have any differences. The satisfaction of the students learning with the application for learning place value was found at the highest level in overall after lesson.*

*KEYWORDS*

*Application, Place Value*

## 1. INTRODUCTION

Mathematics is essential and rudimentary in real-life application and further study, for example, observation, forecasting, logics, and unexpected problem solution efficiently.  To be able to do so, it is called a mathematic proficiency. It required ones to learn from the basic to continue further lessons in the future when it comes to mathematics. Administering teaching and learning plans according to the central fundamental curriculum B.E. 2551 [1] started from mathematic calculation of place value from pratomsuksa 1. Currently, assistive media for lesson revision after a traditional class varied, for example, revision books for students, wooden number media, still figure[2], and audio-visual media[3]. These mentioned media did not build more relationship with students; as a result, it failed to interest students. Therefore, the development was aimed to develop the learning tangible media about addition and subtraction in mathematics in pratomsuksa 1 to practice place value. The digital screen showed number that was equipped with responsive audio, was portable, weighed lightly and input by number buttons [4]. However, the media could not be conveniently portable. The study thereby aimed to develop the application for learning place value for pratomsuksa 1 students via their mobile phones in order to support or revise descriptive lesson about place value, solved portability problems, enhanced sound track and support ubiquitous learning. This was a pathway to develop the students' academic achievement of mathematic place value and interested students while enjoying learning mathematic as a pathway to more efficiency learning media.

## 2. OBJECTIVES

2.1 To develop the application for learning place value.
2.2 To determine the efficiency of developed application
2.3 To compare academic achievement to compare academic achievement between pre-lesson and post-lesson of students who learned with the developed application about place value.
2.4 To compare academic achievement between students who learned place value by the developed application and through traditional method during post-lesson.
2.5 To compare retention of academic achievement of application for learning place value after 2 weeks.
2.6 To find satisfactory of student who learned the application for learning place value.

## 3. OPERATIONAL DEFINITION

Place value defined as a value determining a position of number, for example, decimal place value such as ones digit (1 or $10^o$), tens digit ($10^1$), hundreds digit ($10^2$), thousands digit ($10^3$), ten-thousands digit ($10^4$), hundred-thousands ($10^5$), and millions digit ($10^6$).

## 4. METHODOLOGY

4.1. Population and Sample
4.1.1 Population
4.1.1 Professors through synthesis and quality search of media divided by 2 groups.
4.1.1.1 Specialists evaluating content were specialized at in-depth interview about suitable activities and test quality of teaching mathematics in pratomsuksa 1 with less than 10 years of working.
4.1.1.2 Technical assessment specialists completed Doctoral degree taking an academic position of at least assistant professor in mathematics department, educational media, computation or any related branches or anyone with at least 10 years of teaching experience.
4.1.2 Pratomsuks 1 students of semester 2017 in the central part of Thailand.
4.1.2 Sample Group
4.1.2.1 5 content assessment specialists selected through purposive sampling.
4.1.2.2 5 technical assessment specialists selected through purposive sampling.
4.1.2.3 400 pratomsuksa 1 selected through purposive sampling.

### 4.2 Independent Variables and Dependent Variables

4.2.1 Independent variables defined as the application for learning place value.
4.2.2 Dependent variables defined as efficiency of the application for learning place value, academic achievement, retention of learning, and satisfaction toward the application for learning place value.

### 4.3. Research Procedure

The application for learning place value was designed and developed under these 5 stages based

The International Journal of Multimedia & Its Applications (IJMA) Vol.10, No.1/2/3, June 2018

on ADDIE Model [5] as below.

4.3.1. Analysis: Documents and theory based on academic management of learning plans, purposes, interview with mathematics teachers about various topics such as features of students, expected behaviors of students, limitation of learning and teaching methods and study of designing and developing application.

4.3.2 Design Phase divided into 2 parts as following

4.3.2.1 The application was designed for learning place value based on tangibility of design construction and learning media development for pratumsuksa 1 students [6] for application and further development. The features of the application are explained as in the figure 1.

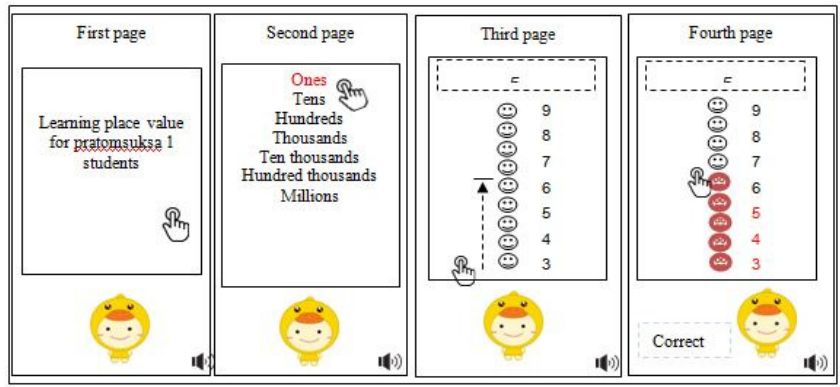

The figure 1. Overall functions of the designed application.

From the figure 1, the first cover consisted of the title, motion cartoon character, narrative audio, and volume on/off.

The second page consisted of number icons starting from ones digit to millions digit. When a mouse touches on each digit icon, narrative audio of main characters turned on. As the ones digit icon was clicked in the second page, the third page appeared consisting of randomized ones digit numbers; for example, a question determining 5 where students had to click counting 5 times until the system showed red number of counting times in order along with the character producing number-count narration upon students' clicks. When the screen showed correct result, the character would say correct. In the case of a wrong answer, the screen showed a retry to the students and retracted to the third page for a retry. If students clicked hundreds digit numbers in the second page of the application, it showed in the figure 2.

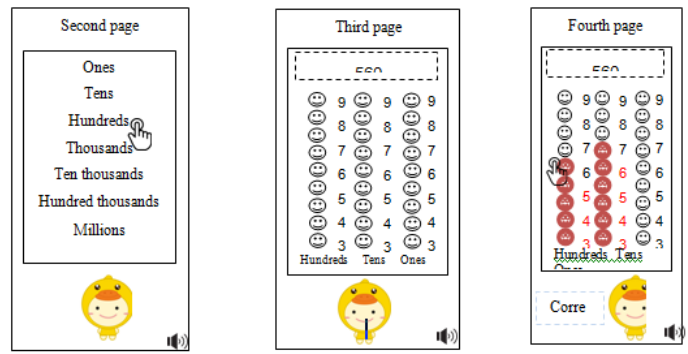

The figure 2: the example of when students click on a hundreds digit icon.

From the figure 2, the question showed a randomized number for students to improve mathematic thinking skilling. The test randomly allowed students to do starting from ones digit to millions digits – 20 questions for each digit and 140 in total during pre and post-lesson. For example, the case of the question randomized 560, it showed how-to-think in fourth page as in the figure 3.

```
560
 ▼
 0   was in ones digit because counting principle did not count zero so clicking it was not
required.
 ▼
 6   was in tens digit allowing students to count tens digit numbers for  6times equal to 60.
 5      was in hundreds digit enabling students to count hundreds digit numbers for  5times
```

According to figure 3, the place value was calculated with the developed application.

4.3.2.2 Content Design: The researcher designed the test for pre - post lesson and during lesson according to the learning plan, teacher guide based on the central fundamental curriculum B.E. 2551 [7] and designed satisfactory questionnaire that was later assessed by 5 content specialists with appropriate content and quality of the questionnaire.

4.3.3 Development Phase: The application was developed centering in Hybrid Apps that was HTML5 CCS3 based and JavaScript (jQuery Mobile Framework) conducted through the co-function of PHP and MySQL to manage the database. This later was designed by 5 technical design and application development specialists to assess the developed application and solved the problems according to their comments as in the figure 4.

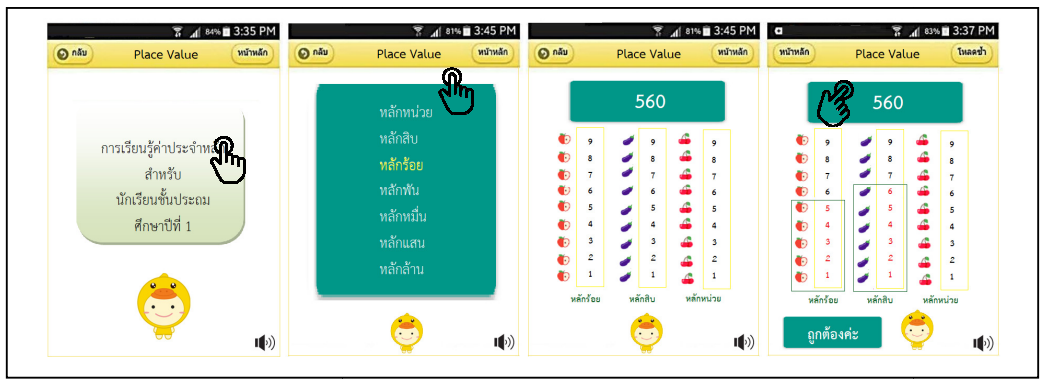

From the figure 4, the sample of the application feature was developed. The students could click to previous lesson for revision or when they wished to select new numbers.

4.4.4. Implementation phase: The application was developed to test with the sample group being 400 students in the central part selected through purposive sampling. The phases consisted as following.

4.4.4.1 Students doing the test during pre-lesson
4.4.4.2 Teachers teaching the application usage and students practicing
place value from the application.
4.4.4.3 Students doing the test during in-class lesson
4.4.4.4 Students' extra knowledge on the application
4.4.4.5 Students doing test at the end of academic subject

4.4.4.6 Students completing satisfaction questionnaire
4.4.4.7 Students doing test during post-lesson test
4.4.4.8 Students doing the test to find retention value after 2 weeks of
post-lesson test.
4.4.8.9  Research instruments

1) The application for learning place value.
2) The questionnaire for the specialists to test the quality in
accordance with the standard of 5 rating scales.
3) The pre lesson test consisting of 60 items, the test during lesson
consisting of 60 questions, the pre lesson test consisting of 60 questions, and retention test consisting 60 question – 240 questions in total.
4) Students' satisfaction questionnaire toward the application for learning
place value that was developed by the 3 rating scale standard [8].

4.4.5. Evaluation Phase

The data were collected and analyzed to find the efficiency of the application
for learning place value, academic achievement of the students, their learning retention, and satisfaction of the students toward the application for learning place value.

Analysis of the data to proceeded as following.

4.4.5.1 The analysis of the quality of the application for learning place value through basic statistics such as percentage, average, and standard deviation.

4.4.5.2 The analysis of the efficacy of the application for learning place value under the standard of 80/80 with the formula of E1/E2.

4.4.5.3 The comparative analysis of efficiency between pre-lesson and post-lesson among student learning by the application for learning place value through basic statistic unit such as percentage, average, standard deviation, and t-test value.

4.4.5.4 The comparative analysis of academic achievement between pre and post lesson among students learning by the application for learning place value and those learning in the traditional method though basic statistic unit such as percentage, average, standard deviation, and t-test value.

4.4.5.5 The comparative analysis of learning retention after lesson by the application for learning place value after 2 weeks through basic statistic unit such as percentage, average, and standard deviation.

4.4.5.6 The analysis of the students' satisfaction toward the application for learning place value through basic statistic unit such as percentage, average, and standard deviation.

## 5. RESULTS

Result of the developed application for learning place value as show in table 1-table 6.

The International Journal of Multimedia & Its Applications (IJMA) Vol.10, No.1/2/3, June 2018

Table 1. Result of the evaluation of appropriateness of the developed application for learning place value.

| Item | $\bar{x}$ | S.D. | Translation |
|---|---|---|---|
| 1. Appropriateness of the background color and fonts. | 4.80 | 0.44 | Highest |
| 2. Aesthetics of the page compositions. | 4.60 | 0.54 | Highest |
| 3. Appropriateness of the animation. | 4.80 | 0.44 | Highest |
| 4. Correctness of speech synthesizer. | 4.80 | 0.44 | Highest |
| 5. Appropriateness of the audio feedback. | 4.60 | 0.54 | Highest |
| 6. Media and content consistency | 4.80 | 0.44 | Highest |
| Total | 4.73 | 0.47 | Highest |

From table 1, the qualitative assessment of the developed application for learning place value showed the overall at the highest level ( $\bar{x}$ =4.73, S.D. =0.47)

Table 2. Measuring the efficiency of the developed application for learning place value

| Tests | Percent |
|---|---|
| During lesson (E1) | 89.07 |
| Posttest (E2) | 91.84 |

From table 2, the efficiency of the developed application for learning place value among pratomsuksa 1 students had overall academic achievement equivalent to 89.07/91.84 that was higher than anticipated as 80/80.

Table 3. comparison of academic achievement before and after lesson.

| Test | S | N | $\bar{x}$ | S.D. | t | sig |
|---|---|---|---|---|---|---|
| Pretest | 60 | 400 | 31.71 | 10.69 | 27.75* | 0.0000 |
| Posttest | 60 | 400 | 53.49 | 5.98 | | |

From table 3, the academic achievement between pre-lesson and post-lesson of students learning by the application for learning place value showed an average of 31.71 before lesson and an average of 53.49 after lesson. The academic achievement between pre-lesson and post-lesson of students learning by developed application was compared to find that the post-lesson was higher than the pre lesson at 0.5 statistical significance.

Table 4. comparison of academic achievement Application and lesson.

| Test | S | N | $\bar{x}$ | S.D. | t | sig |
|---|---|---|---|---|---|---|
| Application | 60 | 400 | 53.49 | 5.98 | -16.71* | 1.0000 |
| Lesson | 60 | 400 | 41.33 | 7.77 | | |

From table 4, the academic achievement of the students with the application for learning place value was an average of 53.49 and an post-lesson average of 41.33. The comparison with the academic achievement of students who learned place value by the developed application and traditional method was different at 0.5 statistical significance. The academic achievement of the students learning with the application for learning place value was higher than those learning with the traditional method.

Table 5. Result of the learning retention of the students learning with the application for learning place value after 2 weeks.

| Test | S | N | $\bar{x}$ | S.D. | t | sig |
|---|---|---|---|---|---|---|
| Pretest | 60 | 400 | 53.49 | 5.98 | 1.16* | 0.1246 |
| After 2 weeks | 60 | 400 | 53.53 | 5.97 | | |

From table 5, Learning retention of the students learning with the application for learning place value after 2 weeks did not have any differences.

Table 6. The study of students satisfaction toward the application for learning place value was evaluated using 3 level rating scale.[10]

| Item | $\bar{x}$ | S.D. | Translation |
|---|---|---|---|
| 1. The Application has clear audio sound. | 2.65 | 0.48 | high |
| 2. The Application has pleasant colors. | 2.92 | 0.27 | high |
| 3. The Application is enjoyable. | 2.97 | 0.16 | high |
| Total | 2.84 | 0.13 | high |

From table 6, the result of the students' satisfaction toward the application for learning place value was found that the students had overall high level ($\bar{x}$=2.84, S.D. 0.13) of learning with application for learning place value.

## 6. CONCLUSION

6.1 The qualitative assessment of the developed application for learning place value showed the overall at the highest level ($\bar{x}$ =4.73, S.D. =0.47)

6.2 The efficiency of the developed application for learning place value among pratomsuksa 1 students had overall academic achievement equivalent to 89.07/91.84 that was higher than anticipated as 80/80.

6.3 The academic achievement between pre-lesson and post-lesson of students learning by the application for learning place value showed an average of 31.71 before lesson and an average of 53.49 after lesson. The academic achievement between pre-lesson and post-lesson of students learning by developed application was compared to find that the post-lesson was higher than the pre lesson at 0.5 statistical significance

6.4 The academic achievement of the students with the application for learning place value was an average of 53.49 and an post-lesson average of 41.33. The comparison with the academic achievement of students who learned place value by the developed application and traditional method was different at 0.5 statistical significance. The academic achievement of the students learning with the application for learning place value was higher than those learning with the traditional method.

6.5 Learning retention of the students learning with the application for learning place value after 2 weeks did not have any differences.

6.6 The result of the students' satisfaction toward the application for learning place value was found that the students had overall high level ($\bar{x}$=2.84, S.D. 0.13) of learning with application for learning place value.

## 7. DISCUSSIONS

7.1 The overall result of qualitative assessment of the developed application for learning place value was at the highest level ($\bar{x}$ =4.73, S.D. =0.47) due to suitable screen, colorful content suitable for the students and its easy-to-use and stable system.

7.2 The efficiency of the developed application for learning place value in pratumsuka 1 student mathematics showed overall result after qualitative assessment equivalent to 89.07/91.84, an average of 89.07 during lesson and an average of 91.84 after lesson. As a result, the developed application had higher efficiency than anticipated as 80/80, meaning it could be used for efficient learning.

7.3 The academic achievement between pre-lesson and post-lesson showed that the students learning with the developed application improved better in the post-lesson than in pre-lesson at 0.5 statistical significance due to that the developed application allowed students to practice learning place value repeatedly and everywhere. For its mobile-supported learning application, student did not have to stay in the class like in the traditional method. Therefore, the academic achievement ws higher than post-lesson.

7.4 The academic achievement of pre-lesson and post-lesson revealed that students learning by the developed application for learning place value had higher achievement than those learning in the traditional method at 0.5 statistical significance. It was because those learning in the traditional method by books and a standstill figure was not interested in learning as the media did not have any interaction with them. They then could not understand and felt bored in the class. However, it was different from the developed application based learning allowing students to develop interaction with finger muscle and eyes [10] while practicing place values until they finally understood and enjoyed learning.

7.5 Learning retention of the students learning with the application for learning place value after 2 weeks did not have any differences due to corresponding with conceptual of Reinforcement theory[11], it must develop media to make students enthusiastic and funny to do more motivation and retention in memory.

7.6 The overall satisfaction of the student toward the application for learning place value was at the high level. The average of the satisfaction toward enjoyment was at the highest due to the instant result of the performances to let the students to know in the developed application. Also, there was a responsive audio to interact with them that raised their learning interests more.